# Energy landscape reveals the underlying mechanism of cancer-adipose conversion with gene network models

Zihao Chen[1,2,#], Jia Lu[3,#], Xing-Ming Zhao[2], Haiyang Yu[3,4,*], Chunhe Li[1,2,5,*]

1 Shanghai Center for Mathematical Sciences, Fudan University, Shanghai, China
2 Institute of Science and Technology for Brain-Inspired Intelligence, Fudan University, Shanghai, China
3 State Key Laboratory of Component-based Chinese Medicine, Tianjin University of Traditional Chinese Medicine, Tianjin, China.
4 Haihe Laboratory of Traditional Chinese Medicine, Tianjin, China.
5 School of Mathematical Sciences, Fudan University, Shanghai, China

[*] Authors to whom correspondence should be addressed: C.L. (chunheli@fudan.edu.cn) and H.Y. (hyyu@tjutcm.edu.cn)
[#] Authors contributed equally: Z.C. and J.L.

**Abstract**

Cancer is a systemic heterogeneous disease involving complex molecular networks. Tumor formation involves epithelial-mesenchymal transition (EMT), which promotes both metastasis and plasticity of cancer cells. Recent experiments proposed that cancer cells can be transformed into adipocytes with combination drugs. However, the underlying mechanisms for how these drugs work from molecular network perspective remain elusive. To reveal the mechanism of cancer-adipose conversion (CAC), we adopt a systems biology approach by combing mathematical modeling and molecular experiments based on the underlying molecular regulatory network. We identified four types of attractors which correspond to epithelial (E), mesenchymal (M), adipose (A) and partial/intermediate EMT (P) cell states on the CAC landscape. Landscape and transition path results illustrate that the intermediate states play critical roles in cancer to adipose transition. Through a landscape control strategy, we identified two new therapeutic strategies for drug combinations to promote CAC. We further verified these predictions by molecular experiments in different cell lines. Our combined computational and experimental approach provides a powerful tool to explore molecular mechanisms for cell fate transitions in cancer networks. Our results revealed the underlying mechanism for intermediate cell states governing the CAC, and identified new potential drug combinations to induce cancer adipogenesis.

**Introduction**

Cancer formation is a complex process with multiple transition states governed by underlying gene regulatory networks (1, 2). Tumor cells undergo both epithelial-mesenchymal transition (EMT) and mesenchymal-epithelial transition (MET) to obtain plasticity and metastasis (3–6). The EMT process enhances the plasticity of the cancer cells, and has been suggested to promote cancer metastasis (7–9). It has been proposed that cells undergoing EMT and/or MET are in a highly plastic state, which may offer a window of opportunity for therapeutic targeting (7, 10, 11). For example, recent work suggested that cancer cells can be induced to adipocytes from certain combination treatments (12, 13), while the inherent mechanisms for cancer-adipose conversion (CAC) have yet to be clarified. This motivates us to explore the underlying molecular regulatory networks controlling the cell fate decisions in CAC.

Dynamical modeling approaches provide effective tools to analyze the functions and behaviors of biological networks, e.g., for EMT and cancer network (14–18). Meanwhile, the stochastic description needs to be considered in cell fate decisions, due to inherently intrinsic and external fluctuations in cells (19, 20). The energy landscape theory, as an extension of the classic Waddington epigenetic landscape metaphor (21), has been developed to study the stochastic dynamics of gene regulatory networks (22–31), e.g., in development and cancerization (24, 32–35). From the landscape view, cell types can be characterized by the basins of attraction on the landscape, which reflect the probability of appearance of different cell types. State with lower potential or higher probability represents attractor or biological functional state, forming the basin of attraction or stable state. So, a biological process such as tumorigenesis or differentiation can be understood as the transition from an attractor state to another one in the gene expression state space of the underlying gene regulatory network (32). Further, the dynamical transition paths between attractors or cell types can be quantified from the landscape based on minimum action path approaches, which provide the information on the order of gene switching on or off in the transition process (34, 36). These approaches provide useful tools to explore the mechanism of the transition from cancer cells to adipose cells.

Here, we seek to unravel the mechanism of CAC using landscape theory. We first constructed a gene network model involving EMT and adipogenesis regulatory network. Based on the CAC model, we quantified the energy landscape to study the stochastic dynamics of this process (22). We identified four types of attractors on the landscape, which characterize epithelial tumor state (E), mesenchymal tumor state (M), adipose state (A) and two partial/intermediate EMT state (P1 and P2), individually. To quantify the transition process in CAC, we calculate the kinetic transition path for each transition. Based on the transition path results, we propose that the CAC can be interpreted as a transition from E or M tumor cell state to A cell state, which agrees well with bulk RNA-seq results (12). To explore the underlying molecular mechanism of drug induced CAC (12), we examine different drug combinations in the CAC gene network model. We find that TGF-β facilitates M state, MEKi promotes the generation of P states with certain level of TGF-β, and Rosiglitazone promotes the A state with TGF-β and MEKi. These results support the hypothesis that malignant tumor cell has the potential of passing through partial EMT state and becoming adipocytes (37). More importantly, the landscape results provide possible theoretical explanations for the mechanism of CAC through intermediate cell states from a molecular regulatory network perspective.

To infer other drug combinations to induce CAC from the landscape model, which requires us to uncover key molecular regulatory elements of CAC, we employ a landscape control approach based on the CAC gene network model. We identify two optimized drug combinations, one of which is the combination of ZEB1 activator with Rosiglitazone, and another one is the combination of MDM2 activator with Rosiglitazone.

To test our predictions for combination drugs and assess their effectiveness experimentally, we established liver, breast, and colon cancer cell lines, applying oil red staining to detect the accumulation of lipid droplets in these cancer cells. Immunofluorescence staining and BODIPY detection by flow cytometry show that tumor cell lipid droplets increase significantly after combined treatment with ZEB1 activator and Rosiglitazone. The Western Blot results show the enhanced expression of adipocyte-related proteins, and RT-PCR results indicate the augment of RNA levels in adipocyte-related genes, after the treatment of combination drugs. Further RNA-seq data analyses show that the cancer cells with the treatment of combination drugs are transformed into cells with more characteristics of mature adipocytes. These experiments verified our model predictions and supported that ZEB1 activator and Rosiglitazone provide a potential effective combination of drug targets for inducing the transition of metastatic tumor cells into adipose cells.

Taken together, our results provide a holistic and quantitative view for CAC and facilitate our mechanistic understanding of cancer to adipose transition process from molecular regulatory network perspective, by combining computational models and molecular experiments. Our work provides a general framework to study the stochasticity and dynamics in cell fate decisions of tumor cells and inspires new ways for cancer treatment.

**Results**

**Construction of a cancer-adipose conversion gene network model**

To explore the molecular mechanism of the CAC process, we first built a gene regulatory network model by combining the EMT and adipogenesis circuits (Fig. 1A, Table S1). Of note, our goal here is not to identify a comprehensive network for CAC, but a core molecular network that can potentially explain diverse cell fate transitions observed in experiments. In fact, recent work suggests that, although regulatory network models may not always include all the regulators involved in cell fate regulations, they still provide incredibly effective tools for understanding cell fate transitions and for making useful predictions (38). To this end, we are focusing on important markers of the EMT network based on previous models (14), in which EMT can be understood as a switching process governed by the reciprocal inhibition between P53-induced microRNAs (miR145, miR200 and miR34) and EMT transcription factors (SNAIL1 and ZEB1) (7, 39, 40). For simplicity, we pick ZEB1 and SNAIL1 to represent ZEB family and SNAIL family, respectively. Moreover, P53 activates its inhibitor MDM2 to form a negative feedback loop (41). The cancer metastasis process involves regulatory loops between P53, microRNAs, OCT4, Let7, LIN28 and BACH1 (42–46). In breast cancer cells, BACH1 promotes the metastasis and inhibits the transcription of the RKIP, while RKIP activates Let7 and inhibits LIN28 and MAPK signaling pathway (18, 47–49).

To acquire a core molecular network of CAC, we integrated the above EMT network with the MAPK signaling pathways (marked by MEK and ERK), the adipogenesis circuit (marked by PPARγ and C/EBPα), and the potential drugs (TGF-β, MEKi and Rosiglitazone) (18, 37, 50, 51). The CAC network involves some important genes related to EMT and adipogenesis such as MEK, ERK, RKIP, SNAIL1, ZEB1, Let7, PPARγ, C/EBPα and corresponding regulations underlying CAC (Fig. 1A). In this network, red arrows represent the activation, and blue bars represent the inhibition (Fig. 1A).

With the network structure, to describe the temporal evolution of different components underlying CAC, we can write down the ordinary differentiation equation (ODE) model (see Materials and Methods and *SI Appendix* for details, see Table S2 and S3 for model parameters, and Table S3 for the robustness analysis of models). The solutions of ODEs characterize multi-dimensional gene expression profiles at different times (Fig. S1). Depending on different initial conditions, the solution of ODEs may converge to different fixed points at steady state, which are also called stable state or attractor in dynamical systems language. It's important to identify multiple attractors in a high-dimensional dynamical system, which correspond to different cell types (Fig. S1 and S2). Also, changes in the parameters of a dynamical system may give rise to bifurcations and alter the number of attractors, leading to a phase transition. Since TGF-β and MEKi have been identified as key factors for CAC in previous experiments (12, 13), we first perform a bifurcation analysis for the model with respect to TGF-β (Fig. 1B-C) and MEKi (Fig. S3).

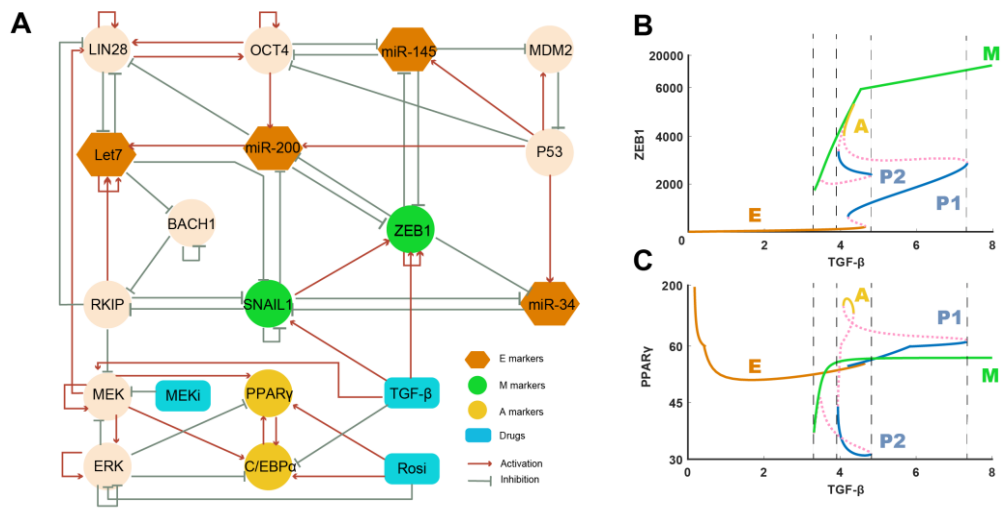

**Fig.1. Deterministic dynamical model for cancer-adipose conversion network**. (A) Wiring diagram of cancer-adipose conversion model. The red arrows represent activation and the blue bars represent inhibition. Circle nodes represent transcription factors, with green nodes representing mesenchymal markers, yellow nodes representing adipogenesis markers. Orange hexagonal nodes represent microRNAs, and blue quadrilateral nodes represent drugs, including TGF-β, MEK inhibitor and Rosiglitazone. (B) Bifurcation analysis of TGF-β with respect to ZEB1. The solid lines with different colors represent the stable states and the pink dotted lines represent unstable states. E represents the epithelial tumor cell state, M represents the mesenchymal tumor cell state, and A represents the adipose cell state. P1 and P2 represent the partial EMT (intermediate) states. (C) Bifurcation analysis of TGF-β with respect to PPARγ.

The bifurcation analysis reveals the appearance of multiple attractors in CAC system with TGF-β raised continuously (Fig. 1B and C). We show the bifurcation diagram for the expression level of mesenchymal marker ZEB1 (Fig. 1B), and the adipogenesis marker PPARγ (Fig. 1C), respectively, with respect to the level of TGF-β. There are up to five stable states (colored solid lines) and three unstable states (pink dotted lines) in the bifurcation diagram. We compare the level of these stable states with experimental data (Fig. 2B), and identify them as epithelial tumor cell state (E), partial EMT cell sate (P1 and P2), mesenchymal tumor cell state (M) and adipose (A) cell state. When the level of TGF-β is very low (<3.2), TGF-β fails to drive the EMT progression and all cells stay in the E state. With increasing level of TGF-β (ranging from 3.2 to 3.9), E and M state coexist in the system, which allows EMT. Further evolution of the attractors with the increase of TGF-β is relatively complex. P2 state arises between the E state and M state with the expression level of TGF-β ranging from 3.9 to 4.8. After generation of P2 state, the A state occurs when TGF-β increases to the range from 4.1 to 4.4. P1 state is between the E state and A state, with TGF-β expression level between 4.3 and 7.3. When TGF-β is in a relative high level (from 4.9 to 7.3), we can observe the coexistence of the M and P1 state in the system. When the level of TGF-β increases extremely high (>7.5), all the cells stay in M state.

Here, a major purpose is to induce the system transition from cancer state (E or M) to A state. Deterministic dynamical model suggests that more than one partial EMT states appear during the CAC. The window of opportunity for the emergence of the A state requires TGF-β level to be maintained within certain range. Furthermore, different partial EMT states provide certain intermediate stations for either M to A transition or E to A transition. With moderate concentration of TGF-β, the transitions of cells from either E or M state to A state pass through the partial EMT states (Fig. 1B and C).

**Energy landscape and transition path quantify the transition process for CAC**

The bifurcation analysis provides initial multistable state information from deterministic point of view. However, it is vital to consider the stochastic dynamics for CAC process since the intracellular noise may play crucial roles in cellular behaviors (20, 52, 53). To disentangle the transition mechanism and study the stochastic dynamics of cancer-adipose conversion, we quantified the corresponding energy landscape based on underlying gene network model of CAC using the approaches we previously developed (Materials and Methods) (22–25, 34). For visualization, we use the dimension reduction approach of landscape to project the landscape onto a two-dimensional space (Fig. 2A, Table S4) (54). On the landscape the blue region represents lower potential or higher probability while the yellow region represents higher potential or lower probability. Consequently, we obtain five attractors on the landscape (Fig. 2A), which correspond to E, M, A, P1 and P2 cell states, respectively (see Table S6 for gene expression levels of five stable states).

To quantify the transition processes between different cell types characterized by attractors on landscape, we employ the minimum action paths (MAPs) method to obtain the transition path for each transition (55). The MAP is the most probable transition path from one attractor to the other. The arrows connecting each cell state represent the MAPs among different attractors. The pink arrows represent progression in the order of E->P1->P2->M, which we assume as the process of EMT. Similarly, the yellow arrows represent the transitions from E or M cell state to the A cell state through P1 or P2 intermediate state. Landscape with five attractors shows that during EMT the epithelial tumor cells are transformed into mesenchymal tumor cells, while the adipogenesis might hijack the EMT toward another committed cell fate, i.e., adipose cell state (37).

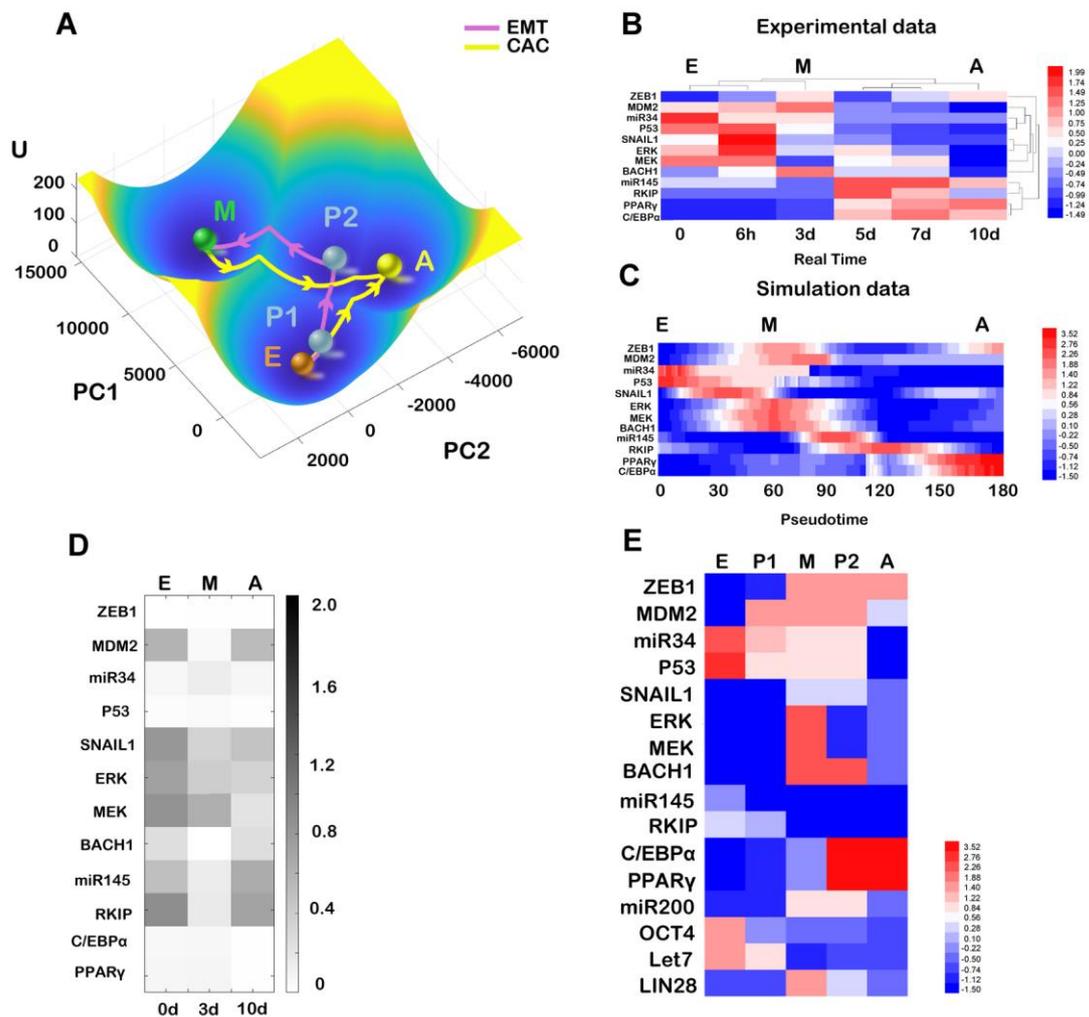

**Fig. 2. Energy landscape and transition path quantify the transition process for CAC.**
(A) The 3-dimensional landscape and corresponding transition paths for cancer-adipose conversion shown in the first and second principal components after dimension reduction of landscape (see Table S4 for how PC1 and PC2 are linearly represented by different genes in CAC network). The yellow lines represent the transition paths of cancer-adipose conversion, and the pink lines represent the transition paths for EMT process. The U axis represents the dimensionless potential. The transition paths represent the most probable path for each transition by minimizing the transition action. (B) Expression levels of different genes in CAC process from bulk RNA-seq data. A clustering method was used to identify the E, M and A phenotypes. By comparing the expression level between model and experimental data, we speculate that the cell state in 7d is P2 state. (C) Pseudotime series expression profiles of different genes for the transition path from E to M and to A from the model. (D) Comparison for the expression level of different genes between model and experimental bulk RNA-seq data. Lighter blocks mean smaller differences between modeling and experimental data. The horizontal axis corresponds to the period of the experiment (d: Day). (E) Expression patterns for five stable states. Red blocks represent high expression levels, and blue blocks represent low expression levels. E represents Epithelial state, M represents Mesenchymal state, A represents adipose state, and P1/P2 represent intermediate states.

Previous studies have documented three distinct states in CAC, corresponding to the E, M and A states (12). Here, we compared the RNA-seq data of mesenchymal breast cells (12) (Fig. 2B) with our pseudotime expression profiles from the model (Fig. 2C) to verify the accuracy of the landscape model and transition paths. Since the experiment begins with the ablation of E-Cadherin gene (12), while the model starts from the E state, we expect that there will be some differences for gene expression level at the initial E state (Fig. 2B and C). Nevertheless, we found the simulation data (Fig. 2C) mostly match the experimental data (Fig. 2B) in terms of the expression of key genes. For instance, the E state shows high expression of p53 and miR34, which are characteristics of the epithelial state (Fig. 2B and C). The expression profiles from both experiment data and modeling have similar M state with high level of ZEB1 and BACH1 (Fig. 2B and C). Regarding the A state, the adipogenesis markers PPARγ and C/EBPα have relatively high expression in both simulation data and experimental data (Fig. 2B and C).

We show the difference in the expression level of each gene for the stable states E, M and A, between our model and bulk RNA-seq data (Fig. 2D). Here lighter blocks mean smaller difference between modeling and experimental data. The results from our model are generally consistent with bulk RNA-seq data (Fig. 2D) (12). We also show the expression patterns of the five attractors (fixed points) on the landscape (Fig. 2E, corresponding to Table S6). From the expression pattern of the five stable states, the E state has the highest expression level of P53, while the M state has the highest expression of MEK, ERK and SNAIL1. P1 and P2 states have lower level of MEK and ERK expression than the M state, and less p53 expression than the E state. P2 and A states both have higher expression levels of adipose markers PPARγ and C/EBPα.

Based on the landscape, we propose two underlying transitions for the A state generation process, i.e., E to P1 and then to A, or M to P2 and then to A. Our quantitative results from landscape and kinetic paths demonstrate that the E state cells will pass through the P1 state to enter A state (Fig. 2A), while the M state cells will pass through the P2 state to enter A state (Fig. 2A). Here, when we employ clustering method on RNA-seq data of the mesenchymal breast cells (12), we identified four clusters, including E, M, A cluster, and another one. Based on the comparisons of gene expressions (Fig. 2B and C), we speculate this additional cluster (experimental data for 7d, Fig. 2B) is the P2 intermediate state, which supports our modeling results on the intermediate states. This conclusion is also supported by the expression profiles based on the multi-dimensional transition paths (Fig. S4 and S5). During the transition from E to A, we identified a region of states which has similar expression level as P1 (by Pearson correlation coefficient, Fig. S4 A and B). We also identified the P2 state along the transition path from M to A, demonstrating that the M->A transition will pass through the P2 state (Fig. S4 C and D). We need to emphasize that, previous work (12) did discuss the possible roles of intermediate states, but no quantitative results or the evidence of existence of intermediate states in CAC are presented. Here we provide the evidence for the existence of intermediate states during the CAC process from a detailed molecular network model.

**MEK inhibitor and Rosiglitazone induce the transition of cancer cells into adipose cell through partial EMT state**

To clarify the stochastic mechanism of cell fate commitment in CAC, we altered the key regulators and track the changes of attractors on the landscape (Fig. 3). To visualize and analyze the high-dimensional landscape, we project the landscape onto a two-dimensional plane to visualize multiple attractors (Fig. 3). Here we choose PPARγ (characterizing the level of adipogenesis) and miR145 (characterizing the level of E state) as the two coordinates. These efforts allow us to understand the CAC from the two coordinates of EMT and adipogenesis. We also show the landscape results using other pairs of genes as the coordinates (Fig. S6). We need to emphasize that our results are based on the full gene network (Fig. 1A), and the landscape is shown in reduced dimensions for the visualization purpose.

To see how the landscape changes with different drug additions, we start from the E state cells without TGF-β addition (Fig. 3A). The E state has a low level of PPARγ but a high level of miR145 (Fig. 3A). We simulate the addition of Rosiglitazone and MEKi but without TGF-β (Fig. 3F). The combination of Rosiglitazone and MEKi increases the PPARγ level, but fails to induce the cell fate transition (Fig. 3F). In another simulation, we add a moderate level of TGF-β (+TGF-β) but without MEKi and Rosiglitazone (Fig. 3B). In this case, we observe the E and M state coexisting (bistable state), which is due to TGF-β inducing the EMT process. In this situation, the E state has a higher level of miR145, while E and M states have similar levels of PPARγ (Fig. 3B). Furthermore, when a high level of TGF-β (++TGF-β) was added, it drives all cells to M state with a lower level of miR145 and PPARγ (Fig. 3C). Of note, the addition of Rosiglitazone and MEKi under high level of TGF-β (++TGF-β) is not able to induce P state or A state (Fig. 3H). These results illustrate that an appropriate range of the TGF-β level is required for inducing CAC. That is because at a low level of TGF-β the system is in a monostable E state (Fig. 3A and F), while at a high level of TGF-β the system is in a monostable M state (Fig. 3C and H). To make the transition to A state possible, we need the system being pushed to a multistable state with A state and intermediate states (Fig. 3G). Therefore, the landscape results with different drug additions (Fig. 3) explain why TGF-β needs to be in a middle range so that the CAC is possible.

Nevertheless, adding TGF-β alone is not enough for the generation of P1 state or A state (Fig. 3A-C). Rosiglitazone and MEKi are also indispensable for CAC. To further clarify the roles of MEKi and Rosiglitazone in the CAC, we fix the concentration of TGF-β to an appropriate level (+TGF-β) and add MEKi and Rosiglitazone individually (Fig. 3D and 3E). When we add Rosiglitazone under moderate level of TGF-β (+TGF-β), the system remains with two stable states (E and M state, Fig. 3D). However, when we add MEKi under moderate level of TGF-β (+TGF-β), the system displays three stable states, including E, M and P1 state (Fig. 3E). The P1 state stays in the middle of E and M state under the EMT coordinate and possesses more PPARγ than M and E state (Fig. 3E). This result demonstrates that the MEKi combined with TGF-β can induce the generation of P1 state.

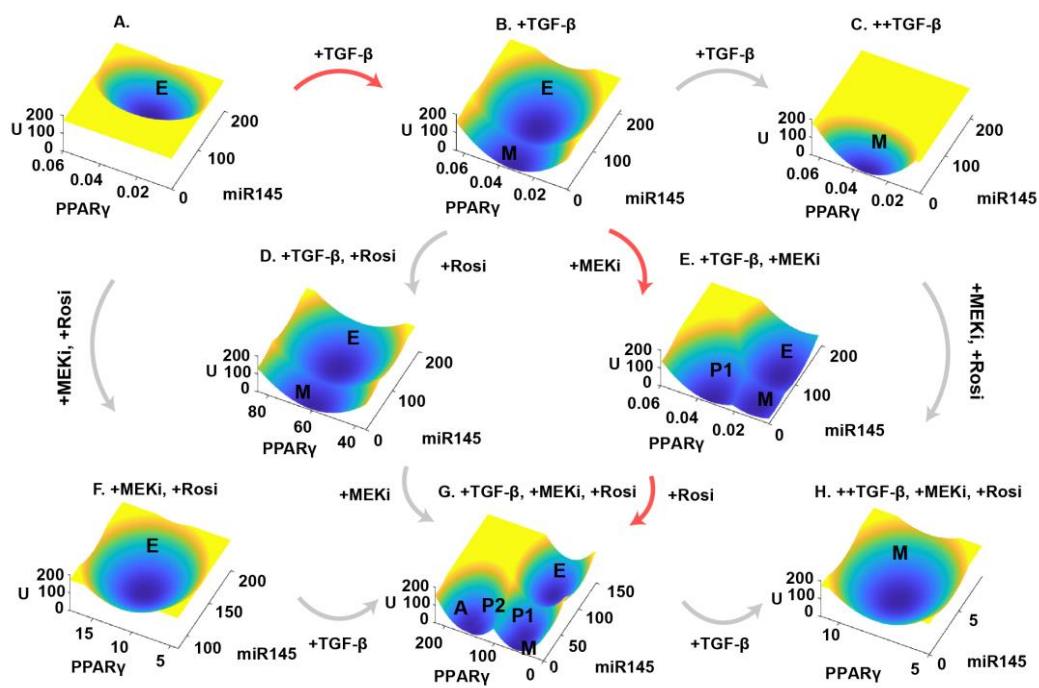

**Fig. 3. Landscapes recapitulate the effects of drug combinations used in previous experiments for inducing CAC.** (A-H) Landscape with different doses and combinations of TGF-β, MEK inhibitor (MEKi) and Rosiglitazone (Rosi). The landscape is projected onto the miR145 axis (characterizing EMT level) and the PPARγ axis (characterizing the level of adipogenesis). The U axis is the dimensionless potential. The red arrows indicate the route for drug additions of inducing CAC proposed in previous experiments (12). Here we use "+" to mark the level of the drugs. (A) TGF-β, MEKi and Rosiglitazone are not added, and the system has a monostable E state, with TGF-β=1, MEKi=0.001, Rosi=0.1. (B) Moderate TGF-β is added (+TGF-β or TGF-β=4.3), but MEKi and Rosiglitazone are not added. (C) High level of TGF-β is added (++TGF-β or TGF-β=8), but MEKi and Rosiglitazone are not added. (D) TGF-β and Rosiglitazone are added (+TGF-β, +Rosi or Rosi=1.5), but MEKi is not added. (E) TGF-β and MEKi are added (+TGF-β, +MEKi or MEKi=0.02), but Rosiglitazone is not added. (F) MEKi and Rosiglitazone are added (+MEKi, +Rosi), but TGF-β is not added. (G) Moderate TGF-β, MEKi and Rosiglitazone are added (+TGF-β, +MEKi, +Rosi). (H) High level of TGF-β, MEKi and Rosiglitazone are added (++TGF-β, +MEKi, +Rosi).

When we add moderate level of TGF-β (+TGF-β) but without MEKi and Rosiglitazone the system keeps in bistable state (Fig. 3B), whereas when we add moderate level of TGF-β (+TGF-β), MEKi, and Rosiglitazone simultaneously, we obtained five stable states (Fig. 3G).

The P2 state has higher level of PPARγ than P1 and M states, while A state has a significantly higher level of PPARγ than other states (Fig. 3G). These results suggest that Rosiglitazone combined with TGF-β and MEKi promotes the adipogenesis by promoting the P2 state and A state (Fig. 3G). During the EMT process, MEKi promotes the potential of the system to generate the intermediate states (or partial EMT states), while the Rosiglitazone facilitates the transition from the intermediate states to adipose state (Fig. S7). Interestingly, our model can recapitulate well the experimental results for combination drugs inducing CAC proposed. Here, the red arrows indicate the dosing procedure proposed in previous studies (12, 13) (Fig. 3). Our results suggest that switching the dosing order between MEKi and Rosiglitazone does not affect the landscape and outcome of CAC (Fig. 3). Therefore, the landscape results from gene network model offer a quantitative explanation for the underlying mechanism of combination drugs inducing CAC.

**Landscape control identifies new drug combinations for inducing CAC**

The landscape with five stable states (Fig. 2A) provides a holistic view for the dynamical process of CAC, i.e., the CAC can be viewed as a transition from E or M attractor to A attractor on the landscape. The CAC process can be also understood as a two-step process illustrated by a cartoon (Fig. 4A). The first step is to induce the appearance of the A state and intermediate (P1 and P2) states. The second step is to increase the occupancy of A state. The balls in the bucket represent the P1, P2 and A state in the EMT process, and the two pieces of woods on the barrel represent the E state and M state. Without the drug treatment, E and M states are the only two stable states and dominate the performance of the system, while the intermediate (P1 and P2) states and A state are hard to appear (Fig. 4A).

Our results resemble the idea that cancer cells can be characterized by the attractors (56), which are determined by underlying regulatory networks. So, therapeutically, a good strategy should be targeting the cancer network to induce cell fate transition from cancer cell attractors to non-malignant cell attractors (14). Targeting the fat metabolism pathway has been shown to be a promising way against cancer in different tumor types (57–59). Several drugs have been suggested to be effective against cancers through MAPK pathway (60). Nevertheless, these drugs focus on killing tumor cells (61), but not changing the landscape of CAC gene networks, such as the stability of E and M states. We reason that traditional strategies for killing tumor cells fail to change the topological structure of the landscape and thus might be ineffective for cancer treatment or lead to cancer relapse (32). We propose that a better way for cancer treatment should be changing landscape topography, i.e., making cancer state less stable, e.g., inducing the transition from cancer cell state to adipose cell state by making perturbations to the underlying cancer gene networks. For example, by performing a single-factor global sensitivity analysis on model parameters, we can uncover the critical elements affecting the relative stability for different attractor states (Fig. S11-S13). Meanwhile, the combination therapy may be more effective than using single drug (12, 62), since targeting a single gene or regulation might be inadequate in changing the landscape of the network, whereas altering multiple genes or regulations can better facilitate this modification (Fig. 3). So, a critical issue is how to identify the optimal combination of drug targets to trigger cell fate transition from cancer cells to non-malignant cells, such as adipose cells.

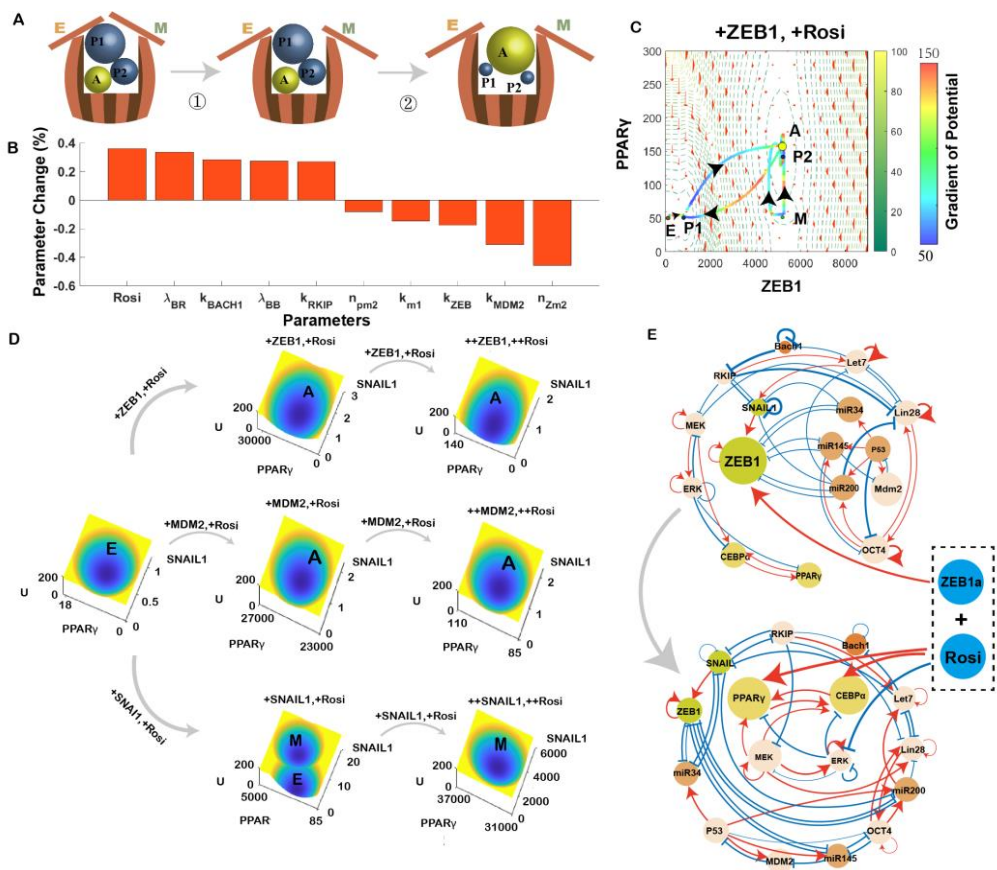

**Fig. 4. Landscape control identifies new drug combination for inducing CAC.** (A) Schematic diagram of the CAC process. (B) Critical elements to increase the occupancy of A state identified from landscape control. The x axis shows the parameters identified from landscape control with the top 10 absolute values of relative change. The y axis shows the relative change ratio of the parameters identified by landscape control with respect to their initial value. (C) Gradient of potential with the addition of ZEB1 activator and Rosiglitazone, with larger balls corresponding to higher occupancy. The color along the paths represents the gradient of potential. The occupancy of A state is enhanced after addition of ZEB1 activator and Rosiglitazone. E, epithelial state; M, mesenchymal state; A, Adipose cell state; P, partial EMT (intermediate) state. (D) Landscape changes for different drug combinations. Gray arrows represent the direction and order of the dosing, with all simulations beginning with no drug additions. The first group of simulated drugs combines ZEB activator and Rosiglitazone with drug level of 10 times (first row, second column) and 500 times (first row, third column). The second group of simulated drugs combines MDM2 activator and Rosiglitazone with drug level of 10 and 100 times (second row). The third group of simulated drugs combines SNAIL1 activator with Rosiglitazone at drug level of 10 and 100 times (third row). (E) Key drug combinations for ZEB1 activator and Rosiglitazone identified from landscape control in terms of CAC network (corresponding to Fig. S8A1 and S8A7). The thickness of the arrows is proportional to the sensitivity of the parameters obtained from the landscape control, and the size of the balls is proportional to the change of the ratio between the synthesis rate and degradation rate for each node.

To enable adipogenesis of malignant cancer cells as a therapeutic option, we aim at reducing the treatment protocol to a minimal number of drug targets in the CAC network. So, the question becomes how to identify the minimal optimal combinations of drug targets to induce cell fate transition from cancer cells to adipose cells, since there will be numerous combinations for drug targets for a large gene network as we studied here (Fig. 1). To this end, we employ a landscape control method to identify the most important regulations to maximize the occupancy of A state (8, 33, 63). The purpose of landscape control is to find a set of parameters which make the objective function obtain the maximum (or minimum) value. Here, we set an objective function based on transition actions to identify the optimal combinations of targets to reach the highest occupancy of A state (Fig. 4B, see Methods for details of landscape control). In the landscape control results, n represents the coefficient of Hill function, and $k_I$ represents the generation rate of the transcription factor I. We performed seven experiments with different initial parameter values (Fig. S8A and S8B) and obtained the average results as the landscape control results (Fig. 4B). Then we picked the top 10 targets (parameters) according to the absolute values of sensitivity (defined as the percentage of the relative change of each parameter at its default vale compared with the value after landscape control) (Fig. 4B). This means that the first five parameters should be increased to promote A state (predicted to be positive change from landscape control), and the latter five parameters should be decreased to promote A state (predicted to be negative change from landscape control) (Fig. 4B). We also predict another potential drug combination for promoting CAC (Fig. S10, Table S5), and illustrate the robustness of the landscape control methods against perturbations on initial parameter values (characterizing heterogeneous tumor populations, Fig. S8).

Since the hill coefficient n, characterizing the cooperativity for regulations, and regulatory strength ($\lambda$) are usually hard to be controlled as a drug target in biological experiments, we pick three major targets from these 10 parameters including Rosi (denoting Rosiglitazone), $k_{ZEB1}$ and $k_{MDM2}$. Therefore, based on the landscape control results (Fig. 4B, 4C and S8), we identify two potential drug combinations, one is activation of ZEB1 and the addition of Rosiglitazone, another is activation of MDM2 and the addition of Rosiglitazone. As an example, we will focus on the drug combinations of ZEB1 and Rosiglitazone. We increased the synthesis rate of ZEB1 and Rosiglitazone to the optimized level identified from the landscape control (Fig. 4C). Here, the size of the ball represents the occupancy of the corresponding attractor states. We find that the occupancy of the A state increases and the occupancy of M state decreases after adding the combination drugs of ZEB1 and Rosiglitazone. Furthermore, all the transition paths are directed towards to the A state, indicating that this combination of drugs could facilitate the CAC process and enhance the transition towards the A state (Fig. 4C). Additionally, the knockdown results of the ZEB1 and Rosiglitazone also illustrate their strong impact on the CAC (Fig. S9). We further observed that at a relatively low level of ZEB1 generation rate and Rosiglitazone addition, the system shifted to a bistable state system (Fig. S9). As previously mentioned (Fig. 4A), the system with only E and M states is unable to generate A state (Fig. S9).

Following the landscape control results, we further test the effects of two kinds of drug combinations from modeling (Fig. 4D). To quantify the efficiency of these drug combinations from modeling, we also provide a control group of drug combinations including EMT transcription factor SNAIL1 and Rosiglitazone, which does not appear in our landscape control results (Fig. 4D). We quantified the drug effects by perturbing the corresponding targets in the network and tracing the attractor modifications. All simulations begin without adding TGF-β, MEKi, Rosiglitazone, or any other drugs, i.e., all the cells start from the E state (Fig. 4D). The first drug combination, i.e., adding the activator of ZEB1 and Rosiglitazone can convert E state to A state successfully, with 10-fold or 500-fold dose of the drugs (Fig. 4D, the first row). The second drug combination, i.e., adding the activator of MDM2 and Rosiglitazone can also transform the cells to A state (Fig. 4D, the second row). The third drug combination, which is the activator of SNAIL1 and Rosiglitazone, was tested as a control example that was not in our prediction (Fig. 4D, the third row). SNAIL1 is another important transcription factor in EMT process. Activation of SNAIL1 and Rosiglitazone with 10-fold dose can induce the transition from the E state to the M state (Fig. 4D). However, as the concentration of the drug combination was increased to 500-fold, high levels of the combination caused all of the cells to transition to the M state (Fig. 4D), without generating adipose cells. In our network, SNAIL1 directly inhibits the RKIP, which is an inhibitor to the MEK pathway. The activation of SNAIL1 might activate the MEK pathway, which goes against the CAC. This might be why ZEB1 can promote the appearance of A state, but not for SNAIL1. These results partially support the effectiveness of combination drugs identified from the landscape control.

As shown on the landscape, the drug combinations predicted from landscape control can generate monostable A states (Fig. 4D, the first and second row), which are better than the combination of TGFβ, MEK inhibitor and Rosiglitazone proposed in previous study (12), which will generate multiple cell types including both adipose cell and other tumor cell states (Fig. 3, red path). This phenomenon can be understood from the network perspective (Fig. 4E). Here, the thickness of the edges represents the weight (corresponding to the sensitivity of parameters) of corresponding regulations, calculated through the landscape control. ZEB1a represents the activator of ZEB1, which intensifies the inhibitory effect on the P53-induced microRNAs (Fig. 4E). This effort drives the EMT process and provides a window of opportunity for the appearance of intermediate states (Fig. 4E). From the network perspective, Rosiglitazone activates the adipogenesis markers C/EBPα and PPARγ which promote adipogenesis. Previous study also suggested that ZEB1 is the core component of adipogenic gene regulatory network, by regulating numerous other transcription factors that promote the development of fat cells (64). These results provide intuitive and quantitative explanations for why the proposed drugs can work for inducing CAC.

**Molecular experiments verified the effectiveness of drug combination for inducing cancer cell adipogenesis**

According to our modeling results, the cancer-adipose conversion can be achieved through drug-mediated EMT process. Next, we aim to perform corresponding molecular experiments to test the effects of the two drug combinations on promoting the CAC. To test the drug effects of overexpressing ZEB1 in combination with Rosiglitazone, we selected relatively more aggressive liver cancer (Hep3B, Huh-7), breast cancer (MDA-MB-231), and colorectal cancer (SW480, SW620) cell lines to construct a ZEB1 overexpression model. When ZEB1 was overexpressed, the cell volume became larger, and the proliferation rate was accelerated. Further, after 24 hours of Rosiglitazone treatment, the cells showed a circular change (Fig. 5A). Colony formation assays showed that overexpression of ZEB1 in Hep3B, MDA-MB-231 cell lines promoted cell proliferation, while further Rosiglitazone treatment for 24h inhibited the cell proliferation (Fig. 5B). Next, we used oil red staining to detect lipid droplet formation in cells. With ZEB1 overexpression and Rosiglitazone treatment, lipid droplet formation can be clearly observed, indicating the possible occurrence of adipose cells (Fig. 5C).

Meanwhile, we conducted western blot and qRT-PCR analyses to detect lipid metabolism and adipogenesis-related markers, including PPARγ, C/EBPα, and Fabp4. The results showed that the combination of ZEB1 overexpression and Rosiglitazone could significantly enhance cellular lipid metabolism and promote fat formation (Fig. 5D). This was also observed in liver cancer cells (Huh-7) and colorectal cancer cells (SW480 and SW620) (Fig. S16 A and S16B). To further verify this, we performed lipid staining with BODIPY, and found a noticeable increase of lipid droplets in tumor cells following combined treatment with ZEB1 overexpression and Rosiglitazone (Fig. 5F). To determine the effect of ZEB1 overexpression and Rosiglitazone combination on tumor cell invasiveness, we used wound healing assay (Fig. 5G) and transwell (Fig. 5H). The results showed that ZEB1 overexpression enhanced tumor invasiveness, while the further addition of Rosiglitazone significantly attenuated the effect of ZEB1 overexpression on tumor cell migration (Fig. 5G and H). Notably, the breast cancer cell line (MDA-MB-231) we used was a P53 mutant cell line, and the liver cancer cell line (Hep3B) was a P53 wildtype cell line. Thus, our combination drugs worked for both cell lines with and without P53 mutation, which was also supported by our modeling results (Fig. S14).

We further tested the effects of another drug combination from our model prediction, i.e., the addition of Rosiglitazone and overexpression of MDM2. We found that MDM2 protein expression was elevated when ZEB1 was overexpressed in Hep3B cell line (Fig. S16C), probably due to the indirect activating role of ZEB1 on MDM2 through miR-145 (Fig. 1A). Similarly, the effects of combination of MDM2 overexpression and addition of Rosiglitazone on promoting CAC were also experimentally verified. As shown in Fig. S16D and S16E, immunofluorescence positive staining for BODIPY and flow cytometry for BODIPY showed that with the combined treatment the tumor cell lipid droplets were increased. At the same time, transwell assay showed that tumor invasiveness was enhanced after overexpression of MDM2, and the addition of Rosiglitazone significantly reduced the ability of the tumor cell migration (Fig. S16F).

To further test the effects of combination drugs (ZEB1 activator and Rosiglitazone) on inducing cancer to adipose transition, we performed bioinformatics analysis to the RNA-seq data from our experiments (*SI Appendix*, Supplemental methods). The results are shown in Fig. S17 (MDA-MB-231 cell line and Hep3B cell line).

For MDA-MB-231 cell line, the ZEB1 overexpression group contains 7198 differentially expressed genes compared to the untreated group, and the group with overexpression of ZEB1 and Rosiglitazone contains 4184 differentially expressed genes compared with the ZEB1 overexpression group (Fig. S17A, top panel). Differentially expressed genes are significantly reduced in the Hep3B cell line (Fig. S17A, bottom panel). We used the KEGG database to further identify the enriched pathways with differential gene expression (FDR < 0.1, p value < 0.05). The results show that when comparing the untreated group and the group with overexpression of ZEB1 and Rosiglitazone, the differential genes are mainly enriched in TNF signaling pathway, p53 signaling pathway, cell cycle-related pathways, and MAPK signaling pathway (Fig. S17B, left panel), as well as apoptosis, glycolysis, and metabolism related pathways (Fig. S17B, right panel). These pathways are all critically related to the hallmarks of cancer (1, 2), illustrating the potential of combination drugs inducing the cell fate transition of tumor cells. We further select several typical pathways (large gene number and small p-value) for the heat map analysis. The results show that genes related to cell migration ability, such as SERPINE1, are significantly downregulated in the group with combination drug treatment, while the cell cycle inhibitory genes CDKN1A, TP53, and PMAIP1 are significantly upregulated in the group with combination drug treatment (Fig. S17C). CEBP, which plays a significant role in adipogenesis, is also upregulated after treated with ZEB1 overexpression and Rosiglitazone. These results support that the treated cells with combination drugs become less aggressive.

To further investigate whether the induced cells are similar to adipose cells, we explored the correlation between different groups after combination drug treatment and the human Simpson-Golabi-Behmel syndrome (SGBS) preadipocyte cell, a typical tool for studies of human adipocytes (GSE161111)(65). Correlation analyses reveal a close resemblance of the differentiated states in SGBS cell line adipogenesis (day 7) with our induced cells after combination drug treatment (correlation coefficient increases from 0.69 to 0.72 for MDA-MB-231 cell line, and correlation coefficient increases from 0.52 to 0.74 for Hep3B cell line, compared with untreated cells, Fig. S17D), which illustrates that the cancer cells with combination drug treatment are induced to cells with more characteristics of mature adipocytes. In summary, the RNA-seq results further support that our predicted combination drugs are effective for inducing the transition from cancer cells to adipose cells.

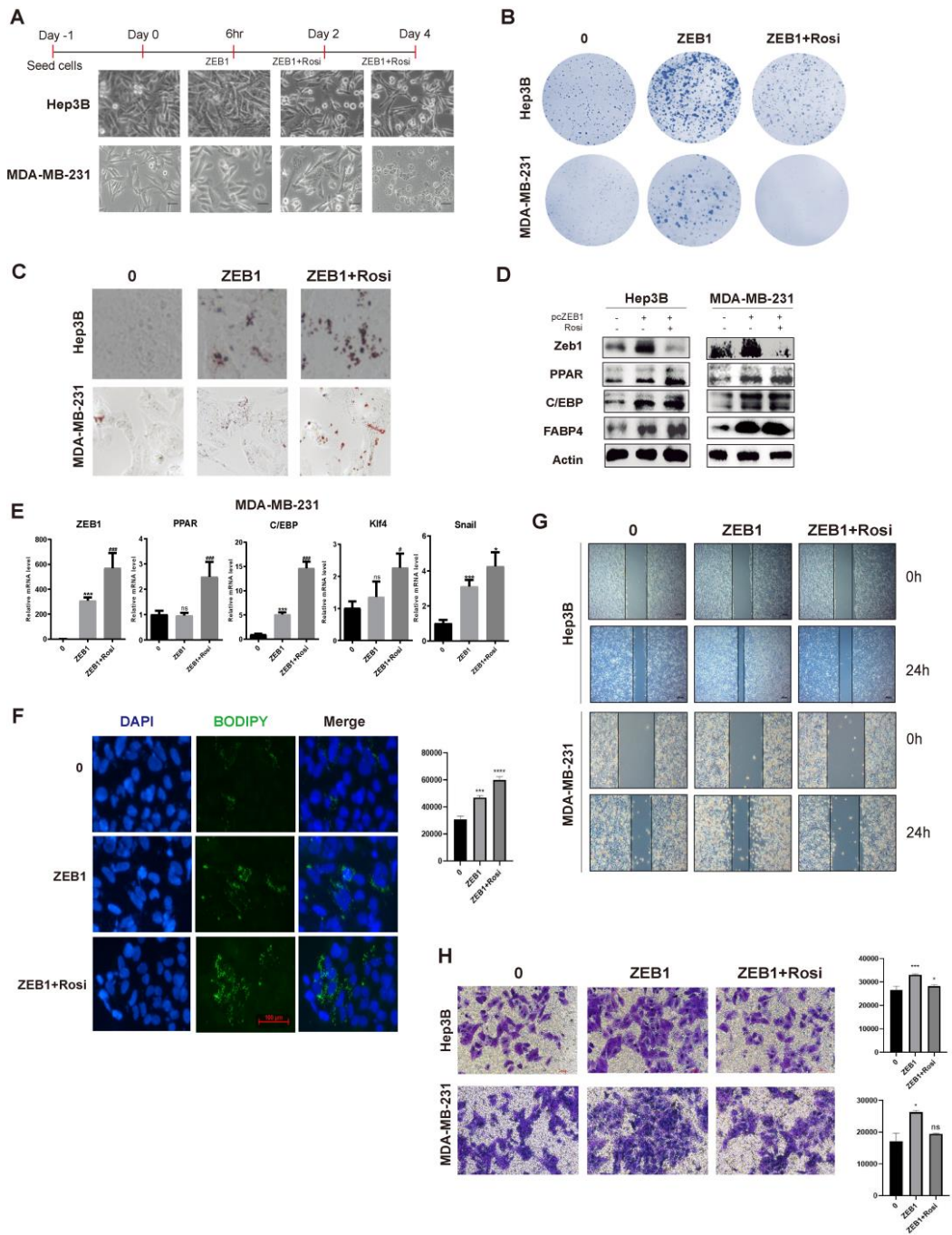

**Fig. 5. The combination of Rosiglitazone and overexpressing ZEB1 promotes the transformation of tumor cells into adipose cells.** (A) Morphological changes of Hep3B and MDA-MB-231 cells exposed to ZEB1 activator and Rosiglitazone for the time indicated (magnification, x100). (B) Colony-formation assays show that ZEB1 overexpression combined with Rosiglitazone reduce the proliferation of Hep3B and MDA-MB-231 cells. (C) Oil red staining was used to detect intracellular lipid droplets in the Hep3B and MDA-MB-231 cell lines (magnification, x100). (D) The expression of adipocyte-related proteins was verified by Western Blot. (E) Changes in related gene expression (RNA) levels such as ZEB1, PPARγ, C/EPBα, Klf4 and Snail were detected by RT-PCR. (F) Bodipy changes in Hep3B cells after ZEB1 overexpression and Rosiglitazone combined treatment were detected by immunofluorescence staining and flow cytometry. (G, H) Overexpressing ZEB1 and addition

of Rosiglitazone reduce the migration of tumor cells as shown in transwell and Wound-healing assays.

**Discussion**

A critical hallmark of EMT in metastasis formation is the acquisition of 'cancer stem cell' properties, which leads to the resistance to therapy (66–69). Traditional cancer drugs which focus on their killing power against tumor cells might be troubled with EMT (70). So, inducing the conversion of cancer cells into adipocytes, which benefits from the EMT, might be a better way in cancer treatment. Previous study proposed that the combinations of TGF-β, MEKi and Rosiglitazone will contribute to the CAC, while the underlying mechanisms have yet to be further clarified (12, 13).

In this study, we constructed a comprehensive gene regulatory network model characterizing the transition process from cancer to adipose cells. To provide a holistic view of the stochasticity and dynamics in the CAC process, we quantified the landscape and transition paths among the five attractors, which correspond to E, A, M state and two intermediate states (P1 and P2). Some studies reported that there were more than one intermediate states during EMT (15, 71, 72), while our results revealed the crucial role of the intermediate states in CAC. Previous experiments observed the transition from M state to A state (12, 13), while our landscape results further revealed two possible transitions, including E->P1->A and M->P2->A (Fig. 2A). This was supported by the pseudotime series expression profiles, which matched the bulk RNA-seq data in the number of clusters and the expression level of regulatory genes. Thus, our analysis clarifies the dynamical mechanism for the transition of cancer cells into adipose cells through the partial EMT (intermediate) states.

To identify the optimized drug combinations for inducing CAC, we employed a landscape control approach, and revealed the key factors to promote the A state. We propose two new drug combinations for promoting CAC, including raising the ZEB1 and Rosiglitazone, and raising the MDM2 and Rosiglitazone. Our molecular experiments further verified the effects of the two combination drugs on promoting CAC (Fig. 5 and Fig. S16).

Previous work suggested a drug combination including TGF-β, MEK inhibitor, and Rosiglitazone for inducing CAC (12, 13). Here, the drug combinations we identified are simpler than theirs in terms of potential for clinical applications, since we only need two drugs for inducing CAC. From modeling results, we also found that the previous combinations with TGF-β, MEK inhibitor, and Rosiglitazone focused on generation of the A state, which failed to make all the stable states associated with tumors become unstable (Fig. 3), while our results provide optimized drug combinations to promote CAC, by eliminating all the tumor associated states and only keep the stable A state (Fig. 4E). Also, our molecular network models revealed the underlying molecular mechanism for CAC regarding the critical roles of intermediate states.

An important insight from the landscape of the CAC is that in the process of transformation from E or M state to A state, cells go through multiple intermediate states. Recent studies have identified intermediate hybrid phenotypes both at single-cell level and population level across different cancer types (7, 72, 73). How P1 and P2 state can be detected explicitly in molecular experiments warrants further explorations. It is important to note that cancer is a complex disease, which involves many hallmarks (1, 2), and this work only modeled a few of these hallmarks (including EMT, metastasis and adipogenesis) using a core molecular network of CAC. Future work can incorporate other critical genes or circuits (and/or other hallmarks of cancer), e.g., important EMT factors TWIST, SNAIL2 and ZEB2, into the models of cancer regulatory networks (14, 74), which may provide more insights into underlying regulatory mechanisms for cancer metastasis and CAC. Our modeling framework also allows for integrating multiple feedback loops between distinct genes or pathways, given the network structure. For example, by integrating the autoregulation of TGF-β into the CAC network, we can obtain consistent results for multiple stable states (Fig. S15).

In this work, our major conclusions focus on the fate transition from cancer cells to adipose cells. We investigate this problem by focusing on the EMT circuit and adipogenesis circuit as well as their interactions. Our experiments have successfully verified the effects of the combination drugs for inducing CAC. However, the tumor metastasis is a very complicated process which involves other molecular regulatory networks. How to regulate the metastasis ability of these adipose cells (e.g., control the dose of drugs) warrants further explorations from both theoretical and experimental efforts.

In summary, our work provides a comprehensive understanding of cancer-adipose conversion through gene network modeling. The landscape and transition paths offer a framework for understanding the underlying mechanisms of cell fate decisions in cancer network and help to design principles to optimize the combination drug strategies for cancer treatment.

**Materials and Methods**

**Cancer-adipose conversion model**

We summarized previous work regarding the regulatory circuits of EMT and adipogenesis (8, 41, 44, 46, 75), and constructed an ODEs model to describe the cancer-adipose conversion. The ODEs share their common forms as in Equation (1):

$$\frac{dX}{dt} = g_X \cdot G - k_X \cdot K \cdot X \tag{1}$$

Here, $X$ represents the level of gene expressions, $g_X$ and $k_X$ represent the basal synthesis and degeneration rate of $X(t)$, individually. $G$ and $K$ denote the regulation of other genes on the synthesis and the degradation of $X(t)$. The regulations among different components can be described by the product of the shifted Hill function: $Hs(Y, S, \lambda, n) = 1 + (\lambda - 1)\frac{Y^n}{S^n + Y^n}$ (14, 76). Here, λ represents the fold change for the regulations, S represents the threshold of the sigmoidal function, and n is the Hill coefficient, which determines the steepness of the sigmoidal function. Y represents the regulator. The Hill function depends on λ in the following way:

$$Hs(Y, S, \lambda, n) \begin{cases} < 1 & \lambda < 1 \\ = 1 & \lambda = 1 \\ > 1 & \lambda > 1 \end{cases} \tag{2}$$

Here, the fold change λ decides whether this regulation is activation or inhibition (λ > 1 for activation and λ < 1 for inhibition) and the strength of regulations. The CAC network comprises a few subnetworks. The first subnetwork is the EMT process, which involves the reciprocal interaction between p53-induced microRNAs and EMT transcription factors (ZEB1, SNAIL1). The second subnetwork is cancer metastasis process, which involves RKIP, Lin28, Let7, Bach1, EMT transcription factors and p53-induced microRNAs. The third subnetwork is MAPK pathway and adipogenesis. MEK is inhibited by ERK, RKIP and other MEK inhibitor (75). ERK is activated by MEK and has both self-activation and self-inhibition effect (77). PPARγ and C/EBPα form a positive feedback loop with each other (78). We also modeled the three drugs corresponding to TGF-β, MEKi and Rosiglitazone by introducing three input nodes in the CAC network (Fig. 1A, see *SI Appendix* for detailed model).

**Self-consistent mean field approximation**

The probability distribution $P(X(1), X(2), ..., X(n), t)$ of a dynamical system is governed by probabilistic diffusion equations, where $X(1), X(2), ..., X(n)$ represent concentration of proteins or gene expression levels in cells. To obtain the probability distribution of a gene regulatory network model, we follow a self-consistent mean field approach (23, 34, 79, 80) to split the probability into products of the individual ones, i.e., $P(X(1), X(2), ..., X(n), t) \sim \prod_i^n P(X_{(i)}, t)$ and solve the probability self-consistently.

Diffusion equations are hard to solve directly for high-dimensional systems, so we started from the moment equations instead. By assuming Gaussian distribution as an approximation, we need to calculate two moments, the mean and the variance. When the diffusion coefficient *D* is small, the moment equations can be approximated by (81, 82):

$$\dot{\bar{\mathbf{x}}}(t) = \mathbf{F}[\bar{\mathbf{x}}(t)] \quad (3)$$

$$\dot{\boldsymbol{\sigma}}(t) = \boldsymbol{\sigma}(t)\mathbf{A}^{\mathbf{T}}(t) + \mathbf{A}(t)\boldsymbol{\sigma}(t) + 2\mathbf{D}[\bar{\mathbf{x}}(t)] \quad (4)$$

Here, $\mathbf{x}$, $\boldsymbol{\sigma}(t)$, and $\mathbf{A}(t)$ are vectors and tensors, and $\mathbf{A}^{\mathbf{T}}(t)$ is the transpose of $\mathbf{A}(t)$. The elements of matrix $A$ are specified as:

$$A_{ij} = \frac{\partial F_i[X(t)]}{\partial x_j(t)} \quad (5)$$

Based on these equations, we can solve $\bar{\mathbf{x}}(t)$ and $\boldsymbol{\sigma}(t)$. Here, we only consider the diagonal elements of $\boldsymbol{\sigma}(t)$ from the mean field approximation. Therefore, the evolution of probability distribution for each variable can be obtained from the Gaussian approximation:

$$P(x, t) = \frac{1}{\sqrt{2\pi\sigma(t)}} exp-\frac{[x - \bar{x}(t)]^2}{2\sigma(t)} \quad (6)$$

Here, $\bar{\mathbf{x}}(t)$ and $\boldsymbol{\sigma}(t)$ are the solutions of Eq. (3) and (4). The probability distribution obtained above corresponds to one stable state. If the system has multiple stable states, there should be several probability distributions localized at each basin with different variances. Thus, the total probability is the weighted sum of all these individual probability distributions. From the mean field approximation, we can extend this formulation to the multidimensional case by assuming that the total probability is the product of each individual probability for each variable. Finally, with the total probability, we can construct the potential landscape by: $U(x)$=-$lnP_{ss}(x)$, with $P_{ss}(x)$ representing steady state probability distribution (22, 23).

**Transition paths and landscape control**

A dynamical system in the fluctuating environments can be addressed by:

$$\dot{\mathbf{x}}(t) = \mathbf{F}[\bar{\mathbf{x}}(t)] + \zeta \tag{7}$$

Here, $\mathbf{x} = (x_1(t), x_2(t), \dots, x_n(t))$ represents the vector of the expression level of proteins or genes. $\mathbf{F}[\mathbf{x}(t)]$ is the vector for the driving force from the dynamical system, $\zeta$ is the Gaussian white noise term, which satisfies $\mathrm{E}[\zeta_i(t)\zeta_j(0)] = 2D\delta_{ij}\delta(t)$. Here, $D$ is the constant diffusion coefficient characterizing the level of noise, $\delta(t)$ is Dirac Delta function, which means that the noises at different times are independent, and $\delta_{ij}$ satisfies:

$$\begin{cases} \delta_{ij} = 1, i = j \\ \delta_{ij} = 0, i \neq j \end{cases} \tag{8}$$

Following the approaches (33, 55) based on the Freidlin-Wentzell theory (83), the most probable transition path from attractor *i* at time 0 to attractor *j* at time T, can be acquired by minimizing the action functional over all possible paths:

$$S_T[\varphi_{ij}] = \frac{1}{2}\int_0^T |\dot{\varphi}_{ij} - F(\varphi_{ij})|^2 \, dt \tag{9}$$

This path is called the minimized action path (MAP). We calculated MAPs numerically by applying minimum action methods (55).

To identify the optimal combination of drugs for promoting CAC, we employ the landscape control method for the CAC model (8, 33, 63). Here, our goal is to predict therapeutic targets (189 parameters characterizing synthesis rate, degradation rate and interaction intensity etc., see *SI Appendix* for details) that can promote the transition from E, M, and partial EMT state to the A state. As such, the optimization process is to minimize the transition action from E, M, and partial EMT state to the A state and maximize the transition action from A to E, M, and partial EMT state (smaller transition action means larger transition probability), by tuning each of 189 parameters. To this end, we define the cost function for maximizing the occupancy of the A state as $\Delta S_A = (S_{M\to A} - S_{A\to M}) + (S_{E\to A} - S_{A\to E}) + (S_{P1\to A} - S_{A\to P1}) + (S_{P2\to A} - S_{A\to P2})$, and our aim is to minimize $\Delta S_A$. We used the Adaptive Minimum Action Method (55) to calculate the transition action, and the matlab function "fmincon" to perform the minimization of transition actions.

**Experimental methods**

**Cell culture**

The breast cancer cell line (MDA-MB-231), hepatoma cell line (Huh-7 and Hep3B), and colon cancer cell line (SW480 and SW620) were obtained from the Cell Bank of the Chinese Academy of Sciences (Shanghai, China), and cultured at 37°C in 5% $CO_2$.

**Western blot**

Cells were lysed in RIPA buffer containing 1 mM PMSF (Solarbio, Beijing, China). The protein concentration was measured using BCA Protein Assay Kit (Thermo Fisher Scientific). Protein samples were boiled and then separated on 8%–10% SDS-PAGE gels, followed by transfer on polyvinylidene fluoride (PVDF) membranes (Millipore). These membranes were blocked for 1 hour in 5% (w/v) skimmed milk at room temperature and then incubated at 4°C with primary antibody overnight. After washing with TBS-T three times, the membranes were incubated with horseradish peroxidase (HRP)-conjugated secondary antibody for 1 hour at room temperature. Finally, the blots were visualized with enhanced chemiluminescence (ECL) reagent (Millipore).

**qRT-PCR**

Total RNA was isolated from cells by using mirVana miRNA Isolation Kit (Life Technologies, Grand Island, NY, USA) or TRIzol Reagent (Life Technologies) according to the standard protocol. For miRNA, reverse transcriptions were performed using the TaqMan miRNA Reverse Transcription Kit (Life Technologies), and cDNA amplification was performed using the TaqMan miRNA Assay Kit (Life Technologies) according to the manufacturer's instructions. The expression of mRNA was determined using the GoTaq qPCR Master Mix (Promega, Madison, WI, USA), with actin used as the endogenous control. Gene expression fold changes were assessed using the 2DCt method. The primers used are listed in following table.

| Pparg2 | GCTGTGAAGTTCAATGCACTGG | GCAGTAGCTGCACGTGCTCTG |
| Klf4 | CGGGAAGGGAGAAGACACT | GAGTTCCTCACGCCAACG |
| C/EBPa | AAACAACGCAACGTGGAGA | GCGGTCATTGTCACTGGTC |
| Zeb1 | GCCAGCAGTCATGATGAAAA | TATCACAATACGGGCAGGTG |
| Snail1 | CTCTGAAGATGCACATCCGAA | GGCTTCTCACCAGTGTGGGT |
| β-actin | TCCCTGGAGAAGAGGCTACGA | AGGAAGGAAGGCTGGAAGAG |

**Wound-healing assay**

Cells were cultured and grown to 90% confluence in 6-well plates and then cultured overnight in serum-free medium. The cell wound was drawn by a 10 mL pipette tip in a straight line. After washing with PBS, wound healing images were taken immediately via an inverted microscope imaging system (Olympus). Then cells were then cultured in medium containing 1% FBS for 24 hours. The 24 hours images were taken in the same way.

**Colony-formation assays**

For a colony-formation assay, 500 cells were seeded in 6-well plates. After incubation at 37℃ about 3 weeks, the colonies were fixed with 4% paraformaldehyde (PFA) for 30 min at room temperature and stained with 0.2% crystal violet for 15 min. The number of visible colonies were counted using Adobe Photoshop (version 2020).

**Oil red O staining**

The cells in each group were washed twice with PBS, fixed with 4% paraformaldehyde for 10 min, stained with oil red O solution for 10 minutes, and then restained with hematoxylin solution for 5 minutes. After washing, under a light microscope, the red granular material in the cytoplasm was the lipid in the cytoplasm stained with oil red O solution, and the nucleus was blue.

**Bodipy dye**

The cells to be stained were removed from the intercellular area and cleaned with PBS before fixation. After fixation, Bodipy working solution was added to stain for 30-60min, and the cell culture plate was wrapped with tinfoil paper to avoid light. After staining, the cells were washed with PBS for 3 times, DAPI working solution was added to stain the nuclei, and PBS was cleaned again for 3 times, and fluorescence microscope was used to take photos.

**Flow cytometric analysis**

Hep3B cells were seeded at a density of $1.5 \times 10^5$ cells/well in six-well plates, grown for 20 hours, and then pre-incubated for 24 hours under ZEB1 or MDM2 with ROSi conditions prior to incubation with BODIPY. After incubation for 30 minutes, the cells were gently scraped, suspended in PBS (Gibco), and transferred to flow cytometry tubes. Subsequently, the cells were analyzed using a flow cytometer (Attune NxT, Thermo Fisher) for Alexa Fluor 488. All analyses were carried out in triplicate using at least 10,000 cells.

**Statistical analysis**

Data are shown as the mean with standard error. The one-way ANOVA or Student's t test was used to determine differences between groups. A p-value < 0.05 was considered statistically significant, and calculations were performed with Statistical Package for Social Science (SPSS for Windows, version 22; Chicago, IL, USA).

**Acknowledgements**

C.L. is supported by the National Key R&D Program of China (2019YFA0709502) and the National Natural Science Foundation of China (12171102). C.L. is a member of LMNS, Fudan University. H.Y. is supported by the Science and Technology Project of Haihe Laboratory of Modern Chinese Medicine (22HHZYSS00008).

**Author Contributions:** C.L. designed research; Z.C. and J.L. performed research; Z.C., J.L., X.Z., H.Y. and C.L. analyzed data; and Z.C., J.L., H.Y. and C.L. wrote the paper.

**Competing Interest Statement:** The authors declare no competing interests.